\documentclass[
reprint,
amsmath,assume,
aps
]{revtex4-2}

\usepackage[dvipsnames]{xcolor}
\usepackage{graphicx} 
\usepackage{subcaption}
\usepackage{derivative}
\usepackage{dcolumn} 
\usepackage{booktabs}
\usepackage{wrapfig}
\usepackage{placeins}
\usepackage{lipsum} 
\usepackage{float}
\usepackage{bm} 
\usepackage{hyperref} 

\usepackage{xcolor}
\usepackage{xparse}
\usepackage{amsfonts}
\usepackage{physics}
\usepackage{parskip} 
\usepackage{etoolbox} 

\AtEndEnvironment{figure}{\ignorespacesafterend}
\AtEndEnvironment{equation}{\ignorespacesafterend}

\usepackage{graphicx}
\usepackage{bm}

\begin{document}

\preprint{APS/123-QED}

\title{Quantum thermalization in a dimerized $J_1-J_2$ model}

\author{Smitarani Mishra}
\author{Shaon Sahoo}
\email{Corresponding author:
shaon@iittp.ac.in}
\affiliation{Department of Physics, Indian Institute of Technology, Tirupati\\
}%

\begin{abstract}
We revisit the $J_1-J_2$ frustrated Heisenberg spin-1/2 chain with dimerization ($\delta$) or modulation in the nearest-neighbor couplings to investigate its thermalization behavior. While the dimerization tends to induce localization, the next-nearest-neighbor interaction $J_2$ generally favors thermalization, making the assessment of the model's compliance with the Eigenstate Thermalization Hypothesis (ETH) particularly subtle. The challenge is further compounded by the model's SU(2) symmetry; the study of ETH compliance is necessarily done for each symmetry sector but separating different sectors of this symmetry is known to be a computationally demanding task. The current study is driven by two main motivations: first, to explore whether the well-known ground-state phases of the model have any bearing on its thermalization properties; and second, to understand how the interplay between two competing factors, namely, the non-uniformity (via $\delta$) and the beyond-nearest-neighbor interactions (via $J_2$) governs the system’s approach to thermal equilibrium.
A systematic analysis shows that the ETH is most strongly satisfied for intermediate values of $\delta$ ($\sim 0.5$) with $J_2$ ranging from intermediate ($\sim 0.5$) to large ($\sim 1$) - a parameter regime falls within the spiral ground-state phase. It is also found that when the system is in the gapless ground-state phase (which falls within the N\'{e}el phase), the ETH is more prone to violation. In the regime of large $\delta$ and small $J_2$, the system is seen to enter a localized phase (characterized here by modulation in density-of-states); assessing ETH compliance is less meaningful for this phase.
\end{abstract}

\maketitle

\section{Introduction}
\label{sec:introduction}

The thermalization of isolated quantum systems remains a central challenge in contemporary theoretical physics, shaped by the complex interplay of quantum coherence, entanglement, and interactions. 
The Eigenstate Thermalization Hypothesis (ETH) has become a key framework for understanding how closed quantum systems thermalize \cite{Deutsch1991, Srednicki1994, Gogolin2016, DAlessio2016, Mori2018}.  The ETH has been extensively supported through studies in a variety of settings, including standard spin models \cite{Alba2015, Steinigeweg2013, Deutsch2018, Dymarsky2018, Mori2018}, disordered spin chains \cite{Suntajs2020}, and models with long-range interactions \cite{Sugimoto2022, Li2016}. However, the ETH breaks down in integrable systems \cite{Rigol2007, Cassidy2011, Wang2024, Sels2021} and many-body localized (MBL) phases \cite{Pal2010, Nandkishore2015, Serbyn2014, Chandran2016}, where usual (Gibbsian) thermalization fails. Its validity has been further examined in disordered Hubbard models \cite{Mondaini2015}, and quantum breakdown scenarios \cite{Lian2023}. Recent experimental advances with programmable quantum simulators \cite{Smith2016, Kaufman2016, Roushan2017} have directly tested ETH predictions, reinforcing its foundational role in quantum statistical mechanics. 

The thermalization in clean one-dimensional spin chains is relatively well understood \cite{Beugeling2014,Noh2021, DAlessio2016, Mori2018}. The presence of disorder, inhomogeneity, or long-range interactions introduces new complexity \cite{Laflorencie2016, Foss-Feig2017, Pino2014, Wang2024, Neyenhuis2017, Schneider2021, Sugimoto2022, Ranabhat2024}. For example, long-range interactions can be shown to be responsible for phenomena such as prethermalization \cite{Neyenhuis2017} and disorder or inhomogeneity may give rise to many-body localization (MBL) and ergodicity breaking (violation of ETH) \cite{Kjall2014,Wang2024}. The dimerized $J_1-J_2$ Heisenberg spin-1/2 chain is a well-known model with a rich ground-state phase diagram \cite{Chitra1995}. 
This model offers a minimal yet nontrivial setting to study thermalization under both inhomogeneity (via dimerization $\delta$ in the nearest-neighbor couplings) and extended-range interactions (via the next nearest-neighbor couplings $J_2$). This, in fact, is a key motivation behind our investigation of the thermalization properties of this model. The other motivation for our current study is to investigate whether the well-known ground-state phases of the model have any connection to its mid-spectrum thermalization behavior.           

To study the thermalization behavior of the spin model and assess the extent to which the ETH is obeyed, we primarily compute the nearest-neighbor spin-spin correlation function $\langle S_1^zS_2^z\rangle$ across all eigenstates of the model. The mid-spectrum fluctuation $\sigma_{sc}$ of this function provides a quantitative measure of ETH compliance, with smaller values indicating stronger adherence. To better understand the thermalization property of the model, we also calculate the effective dimension of subsystem $d_{\text{eff}}$ (a convenient measure of entanglement entropy) and the average von Neumann (VN) entropy $\overline{S}_{vn}$ over all eigenstates.
Our main findings are summarized as follows. (1) When the system has a gapless ground-state phase, the ETH is more prone to violation. (2) In general, the ETH holds more strongly when the system has the gapped spiral phase compared to when it has the gapped N\'{e}el phase. (3) In the regime of large dimerization ($\delta$) and small next-nearest-neighbor coupling ($J_2$), the system enters in to a localization phase; here the energy spectrum exhibits a banded structure with gaps (revealed as a modulation in the density-of-states, DOS). (4) In general, the entanglement entropy across the energy spectrum displays multiple branches due to SU(2) symmetry and localization (connected to the modulation in DOS).   

The paper is organized as follows. We discuss the $J_1-J_2$ model with dimerization in Sec. \ref{sec2}, followed by a discussion on its ground-state phase diagram in Sec. \ref{sec3}. Next, in Sec. \ref{sec4}, we discuss in detail the thermalization behavior of the model in the $\delta-J_2$ parameter space. We conclude our work in Sec. \ref{sec5}.  

\section{Model} \label{sec2}
\begin{figure}[h]
    \centering
    \includegraphics[width=7.0cm, height=3cm]{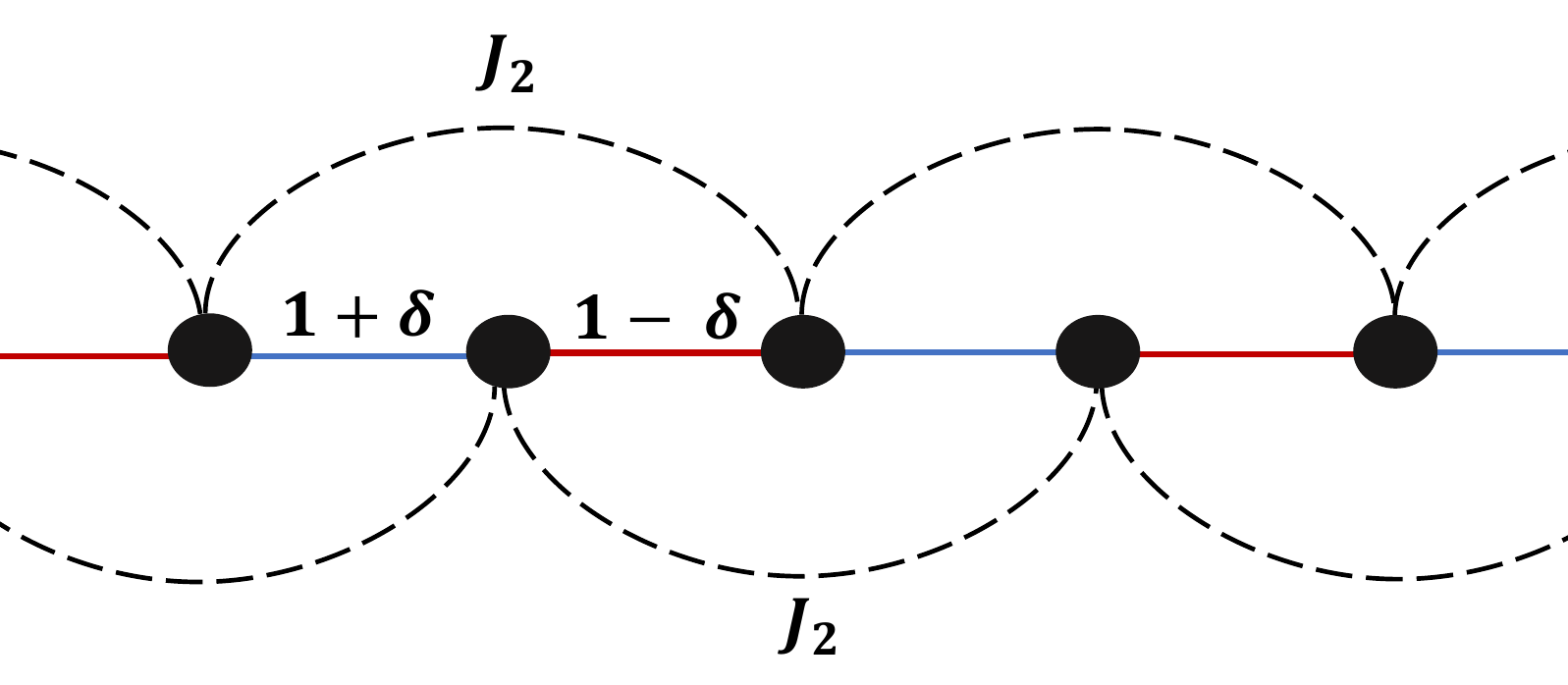}
    \caption{Schematic representation of the \( J_1 - J_2 \) Heisenberg spin chain with dimerization ($\delta$) in the nearest-neighbor interactions ($J_1$). Each filled circle represents one spin.}
    \label{fig:schematic}
\end{figure}

The \( J_1 - J_2 \) Heisenberg spin-1/2 chain with dimerization (staggered nearest-neighbor interactions) \cite{Majumdar1969,Okamoto1992} serves as a prototypical model for studying quantum magnetism with frustration. Its Hamiltonian is given by:
\begin{equation}
H = J_1 \sum_{i=1}^{N}\left[ 1 + (-1)^i \delta \right] \mathbf{S}_i \cdot \mathbf{S}_{i+1} + J_2 \sum_{i=1}^{N} \mathbf{S}_i \cdot \mathbf{S}_{i+2},
\label{eq:hamiltonian}
\end{equation}
where \( \mathbf{S}_i = (S_i^x, S_i^y, S_i^z) \) represents the spin-1/2 (vector) operator at the site \( i \), \( N \) is the total number of sites, and \( \delta \) is the dimerization strength. The parameter \( J_2 \) represents the next-nearest-neighbor exchange coupling, which is responsible for inducing frustration in the system. We take $J_1=1$ for this work. The term \( \left[ 1 + (-1)^i \delta \right] \) introduces a spatial modulation in the nearest-neighbor interactions, leading to a staggered pattern of coupling strengths. The periodic boundary conditions are assumed, such that \( \mathbf{S}_{N+1} = \mathbf{S}_1 \) and \( \mathbf{S}_{N+2} = \mathbf{S}_2 \). The periodic boundary conditions are crucial for avoiding edge effects in finite systems and have been extensively used in numerical studies of one-dimensional quantum magnets \cite{Schollwock2011}. A schematic diagram of the model is presented in Fig. \ref{fig:schematic}.

When \( J_2 = 0 \) and $\delta=0$, the model reduces to a standard nearest-neighbor Heisenberg spin chain. As \( J_2 \) increases, the competition between the nearest-neighbor and next-nearest-neighbor interactions leads to different ground-state phases. A particularly interesting case arises when \( J_2 = 0.5 \) and \( \delta = 0 \), which corresponds to the Majumdar-Ghosh model \citep{Majumdar1969}. In this case, the Hamiltonian becomes:
\begin{equation}
H_{\text{MG}} = \sum_{i=1}^{N} \mathbf{S}_i \cdot \mathbf{S}_{i+1} + 0.5 \sum_{i=1}^{N} \mathbf{S}_i \cdot \mathbf{S}_{i+2}.
\label{eq:majumdar_ghosh}
\end{equation}
The ground-state of the Majumdar-Ghosh model can be found exactly; the ground-state is found to be doubly degenerate, consisting of dimerized spin-singlet pairs. This model provides a key example of frustration in one-dimensional spin systems and has been extensively studied in the context of quantum phase transitions and entanglement properties \citep{Majumdar1969, Shastry1988, Goli2013}.

Inclusion of the dimerization term \( \delta \) further enriches the behavior of the system. The ground-state phase diagram in the $\delta-J_2$ plane is discussed in the next section. 

\section{Ground-state Phase Diagram} \label{sec3}

\begin{figure}[ht!]
    \centering
    \includegraphics[width=7.4cm,height=7.4cm]{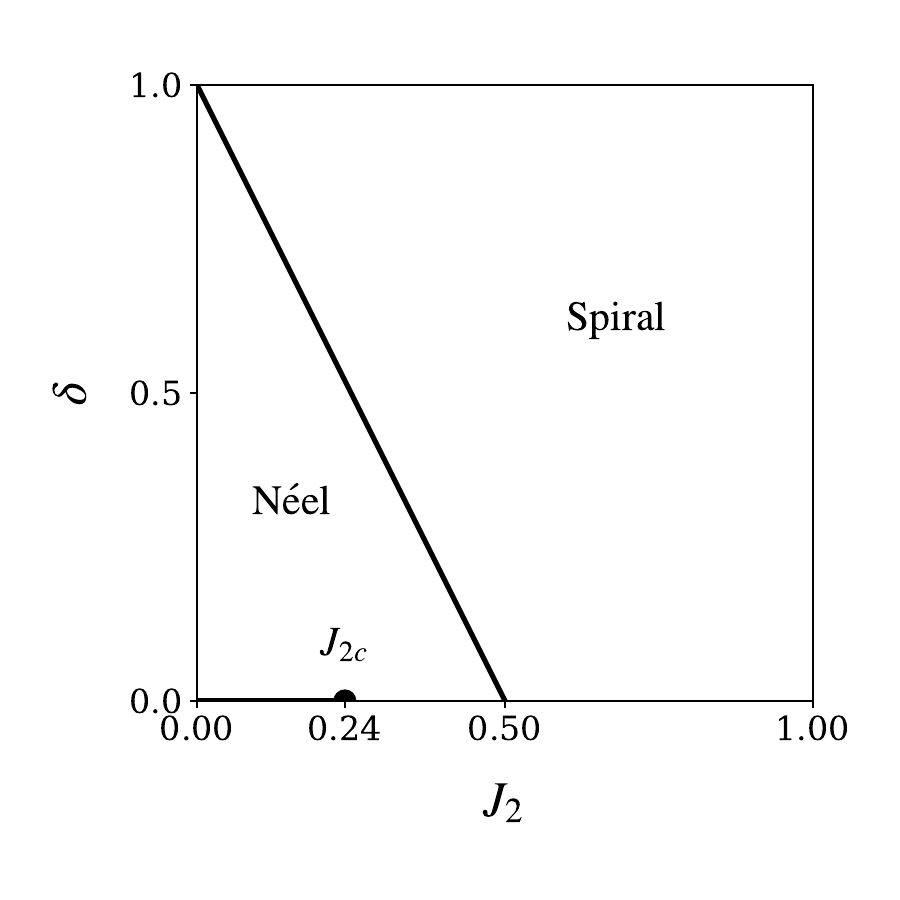} 
    \caption{Schematic ground-state phase diagram of the $J_1-J_2$ spin-1/2 chain with dimerization ($\delta$). The model exhibit two phases - N\'{e}el and Spiral - separated by the straight line $2J_2 + \delta = 1$. Along this line, product of dimers represents the exact ground-state. The whole diagram corresponds to gapped phases, except on the line segment $0\le J_2 \le J_{2c}\approx0.241$ with $\delta=0$ where the system is gapless.}
    \label{fig:phase_diagram}
\end{figure}

In this section, we briefly discuss the ground-state phase diagram of the model \cite{Chitra1995}, which is schematically shown in Fig. \ref{fig:phase_diagram}. 

\begin{figure*}
    \includegraphics[width=16.2cm, height=6.0cm]{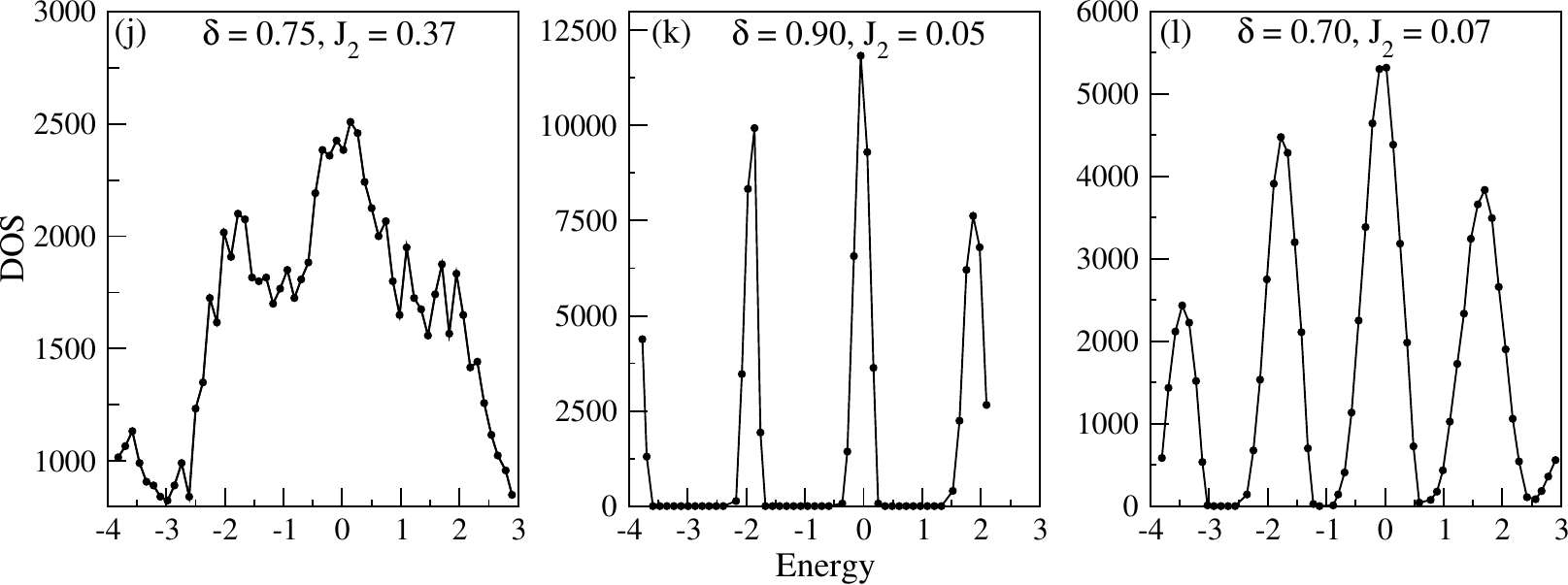}
    \caption{Density of states (DOS) profiles across the energy spectrum for the three selected points (j) - (l) (see Fig. \ref{phase_point}).}
    \label{compare_dos}
\end{figure*}

The uniform chain ($\delta=0$) without frustration ($J_2=0$) is just the Haldane spin-1/2 chain which is known to have gapless ground-state \cite{Haldane1983}. The uniform chain remains gapless with the frustration strength up to $J_2\le J_{2c}\approx 0.241$, beyond which ($J_2> J_{2c}$), the system becomes gapped \cite{Okamoto1992}. The system goes through a quantum phase transition at $J_2=J_{2c}$. In the phase diagram, $J_2=0.5$ represents a special point where the product of the nearest-neighbor dimers represents the exact ground-state of the model (it is also two-fold degenerate) \cite{Majumdar1969}. 

With dimerization $\delta>0$, the ground-state phase becomes gapped \cite{Chitra1995}. In fact, the whole ground-state phase diagram is gapped except when $\delta=0$ and $0\le J_2 \le J_{2c}$. The phase diagram shows a crossover (disorder) line, $2J_2 + \delta = 1$; to the left side of this line, the system shows a N\'{e}eel phase, whereas, to the right side of the line, the system shows a spiral phase. In the phase diagram, another interesting line is $\delta=1$; this line corresponds to a ladder system (two coupled chains) for any finite value of $J_2$.  

\section{Thermalization properties} \label{sec4}
In the previous section, we outlined the ground-state phases of the dimerized \( J_1\text{-}J_2 \) spin-\( \frac{1}{2} \) chain. We now turn to a systematic study of the model’s thermalization behavior. Our analysis is based on exact diagonalization performed within the \( S_z = 0 \) sector, where \( S_z \) denotes the \( z \)-component of the total spin. Although the model possesses SU(2) symmetry - implying that the \( S_z = 0 \) sector can be further decomposed into subspaces with definite total spin \( S \) - numerically resolving these finer sectors becomes increasingly difficult as system size grows~\cite{Sahoo2008, Sahoo2012}. Therefore, we restrict our computations to the full \( S_z = 0 \) sector.

An important consequence of not fully resolving all symmetry sectors is the emergence of multiple branches in plots of the entanglement entropy across the many-body energy spectrum. We return to this point and discuss it in detail later. 

\begin{figure}[ht!]
    \centering
    \includegraphics[width=8.0cm, height=6cm]{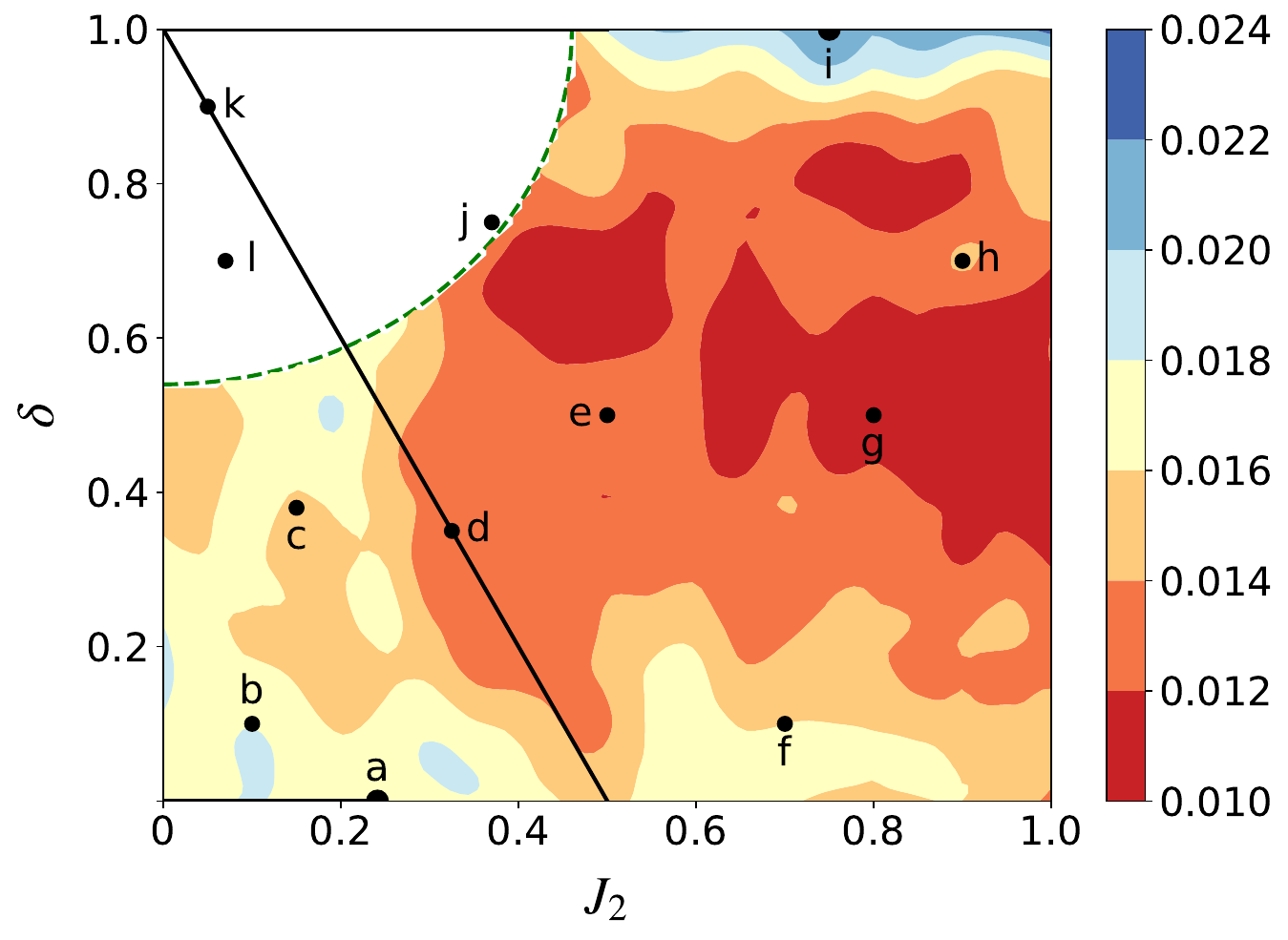}
    \caption{Thermal phase diagram of the \( J_1 - J_2 \) model with dimerization. It shows the value of mid-spectrum fluctuation $\sigma_{sc}$ of a local observable. The disorder line, \(2J_2 + \delta = 1\), is shown on the diagram. The fluctuation is not calculated in the white region which is separated by a dotted line encircling the point ($\delta=1,J_2=0$). 
    The points mentioned on the diagram, ($a$) - ($l$), are separately studied in detail.}
    \label{phase_point}
\end{figure}

\begin{figure*}
    \centering
    \includegraphics[width=14.9cm, height=10.7cm]{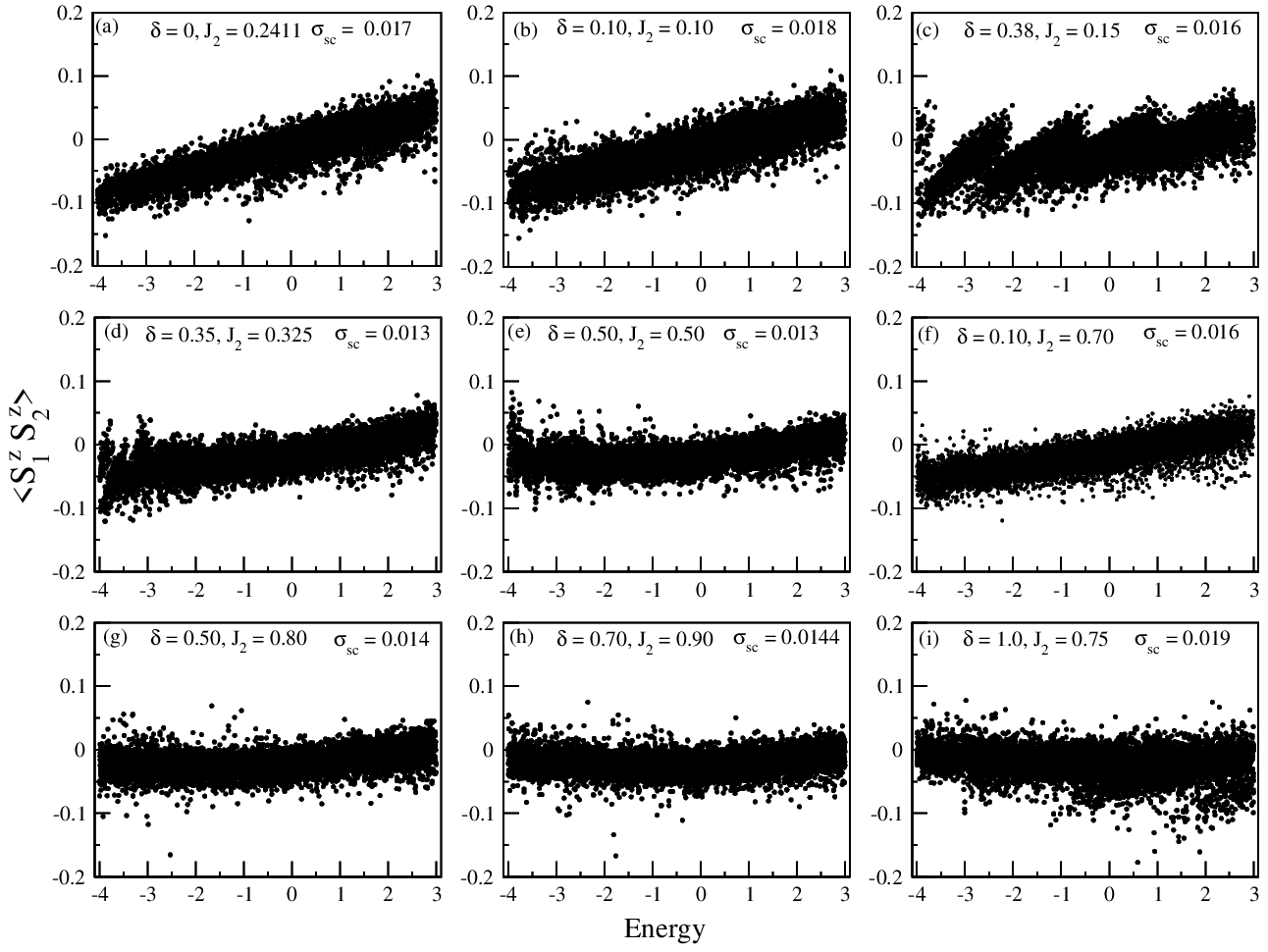} 
    \caption{Expectation value of the local operator for individual eigenstates, $\langle S_1^zS_2^z \rangle$, is plotted across the energy spectrum. Plots are shown for the points ($a$) - ($i$) from Fig. \ref{phase_point}.}
    \label{sz1sz2_exp}
\end{figure*}

\subsection{Local observable and mid-spectrum fluctuation}
Quantum thermalization refers to the process by which, after a long time of evolution, the expectation value of a local observable matches what is predicted by the microcanonical ensemble. This requires that, within a narrow energy window, the expectation values of the observable in different eigenstates be nearly identical. This forms the basis of the Eigenstate Thermalization Hypothesis (ETH), which is typically satisfied in nonintegrable systems \cite{Mori2018, DAlessio2016, Deutsch1991, Srednicki1994}.

The above discussion highlights that fluctuations in the expectation values of a local observable corresponding to eigenstates associated with a narrow energy window serve as a key indicator of ETH compliance — smaller fluctuations imply stronger thermalization and closer adherence to the hypothesis. We take the following operator as the local observable: $O=S_1^zS_2^z$. The expectation value of $O$ gives the nearest-neighbor spin-spin corelation. In the middle of the spectrum (where the density of states is largest), we calculate the fluctuation of the observable in the following way:
\begin{equation} \label{fluct_obs}
\sigma_{sc}=\sqrt{{\overline{O^2}-\overline{O}^2}}.
\end{equation}
Here $\overline{O}$ is the expectation value averaged over different eigenstates, i.e., $\overline{O}=\frac{1}{D}\sum_{i=1}^D\bra{i}O\ket{i}$, where the eigenkets $\ket{i}$ are from the narrow energy window and there are $D$ eigenkets in the window. We can also define $\overline{O^2}$ similarly. 

For system size $N=16$ and subsystem size $N_s=6$, we calculate $\sigma_{sc}$ across the parameter space $0\le \delta \le 1$ and $0\le J_2 \le 1$, except for a region in the top-left corner of the thermal phase diagram (see Fig. \ref{phase_point}). 
In this white region, the energy spectrum exhibits a banded structure with gaps, as is evident from the density of states (DOS) plots in Fig. \ref{compare_dos}. It is not difficult to see why the spectrum behaves this way. When $\delta=1$, the system becomes a ladder; here $J_2$ is the exchange constant along the two legs and $J_1=2$ is the exchange constant along the rungs. Now, when $J_2=0$, we get disconnected rungs. For the collection of disconnected rungs, the spectrum is a set of discrete energies (with degeneracies). As $J_2$ increases, we get a band-like structure that is observed in Fig. \ref{compare_dos}. In this regime (large $\delta$ and small $J_2$), the system is evidently in a localized phase since it corresponds to weakly interacting dimmers (rungs). 

Our numerical results presented in Fig. \ref{phase_point} show that the ETH holds more robustly when the value of $\delta$ is intermediate ($\sim 0.5$) and the value of $J_2$ is in the range between intermediate ($\sim 0.5$) and large ($\sim 1$). In this regime, the fluctuation of the local observable $\sigma_{sc}$ is small, and this region falls within the spiral ground-state phase mentioned earlier. It is interesting to note that, although dimerization induces localization, a delicate balance between finite $\delta$ and $J_2$ here helps the system to thermalize strongly according to the ETH.

For better understanding of the thermalization of the model and the degree of the ETH compliance, for some specific points from Fig. \ref{phase_point}, we plot the expectation values of the local operator  $O= S_1^zS_2^z$ for different eigenstates across the energy spectrum. The results are presented in Fig. \ref{sz1sz2_exp}.

\begin{figure}
    \centering
    \includegraphics[width=8.2cm, height=7cm]{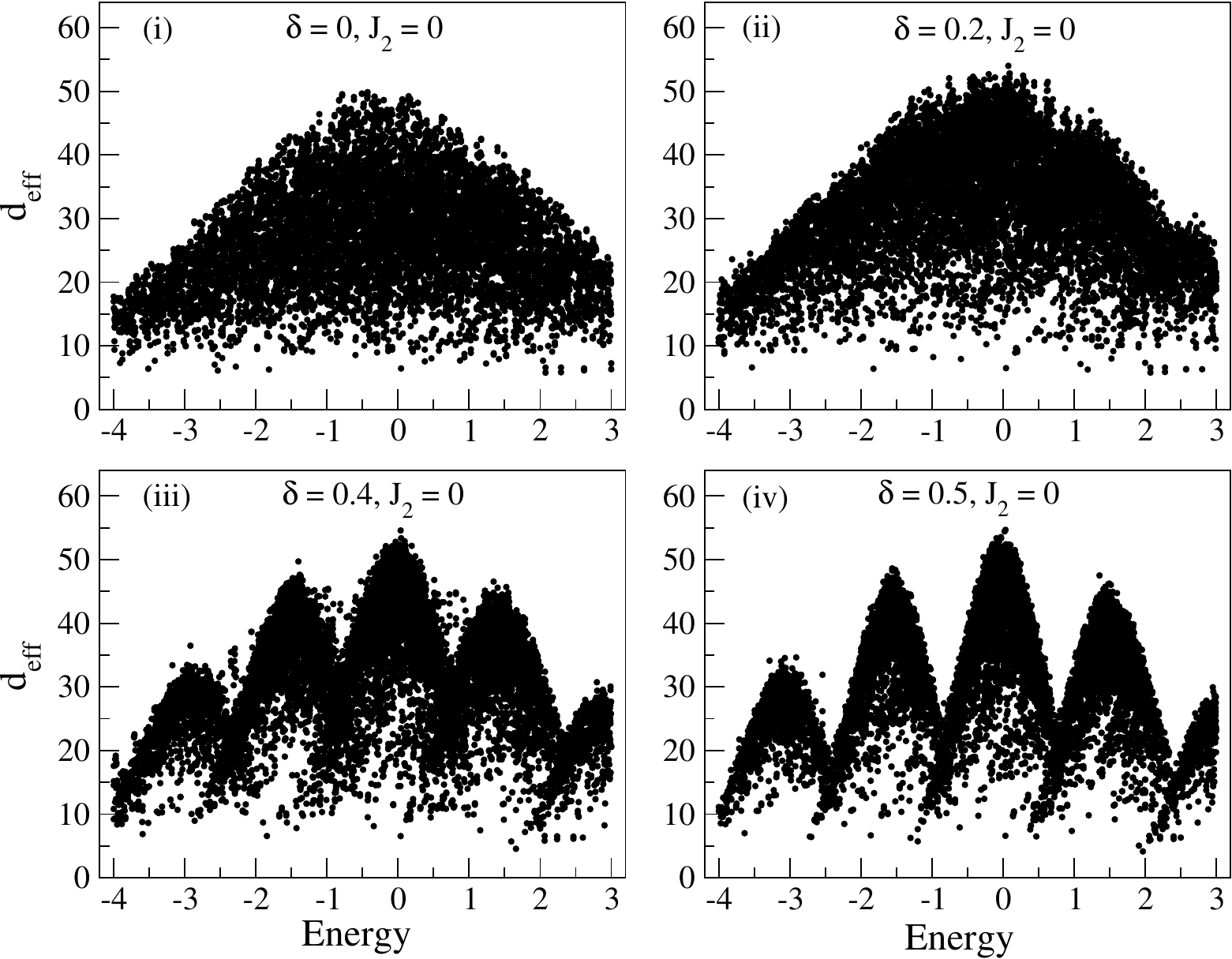}
    \caption{Effective dimension \( d_{\text{eff}} \) of the subsystem for the individual eigenstates across the energy spectrum is shown for some $\delta$ values (with $J_2=0$).}
    \label{deff_j2_0}
\end{figure}

\begin{figure}[t]
    \centering
    \includegraphics[width=8.2cm, height=7cm]{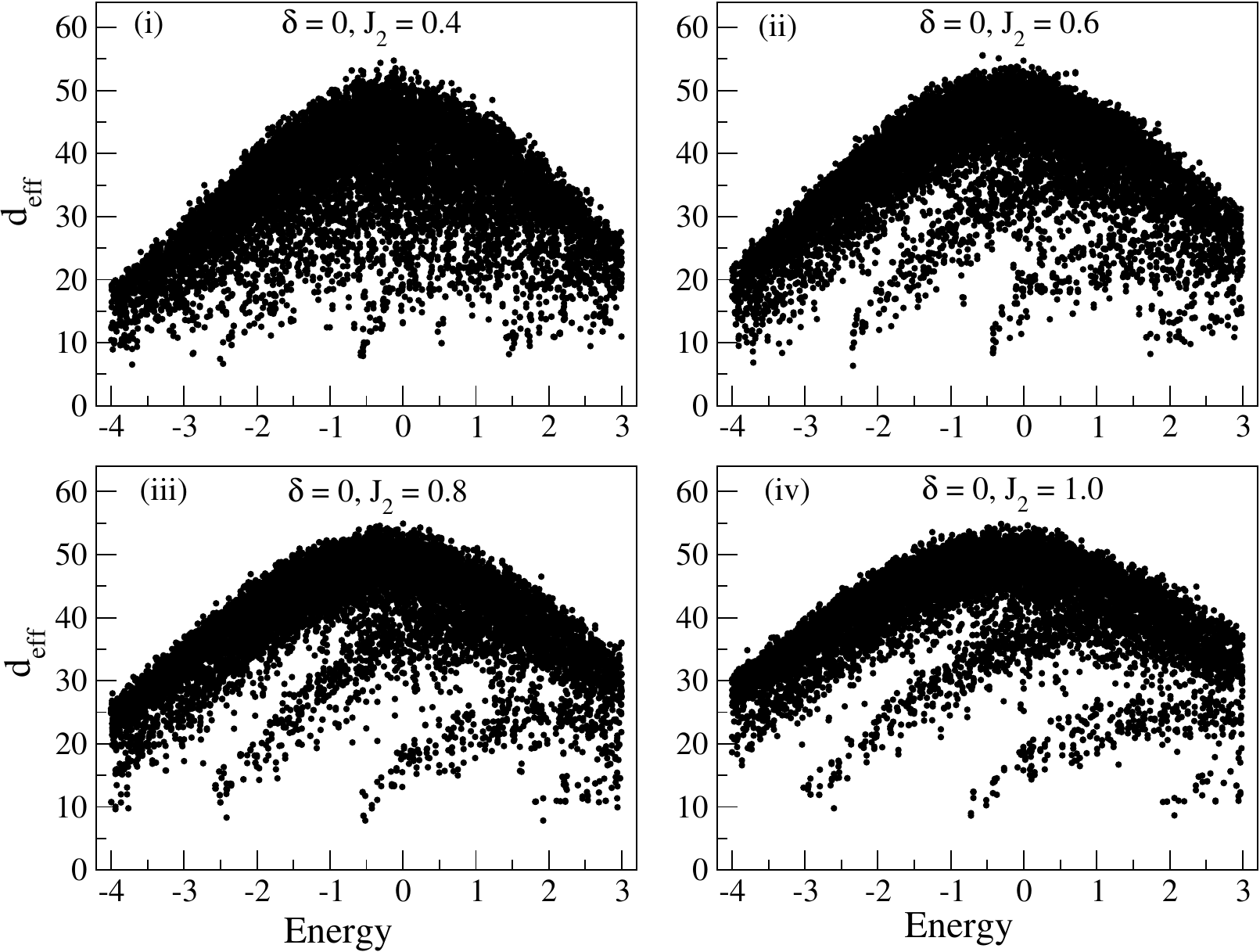}
    \caption{Effective dimension \( d_{\text{eff}} \) of the subsystem for the individual eigenstates across the energy spectrum is shown for some $J_2$ values (with $\delta=0$).}
    \label{deff_del_0}
\end{figure}

\begin{figure}[t]
    \centering
    \includegraphics[width=8.2cm, height=4.0cm]{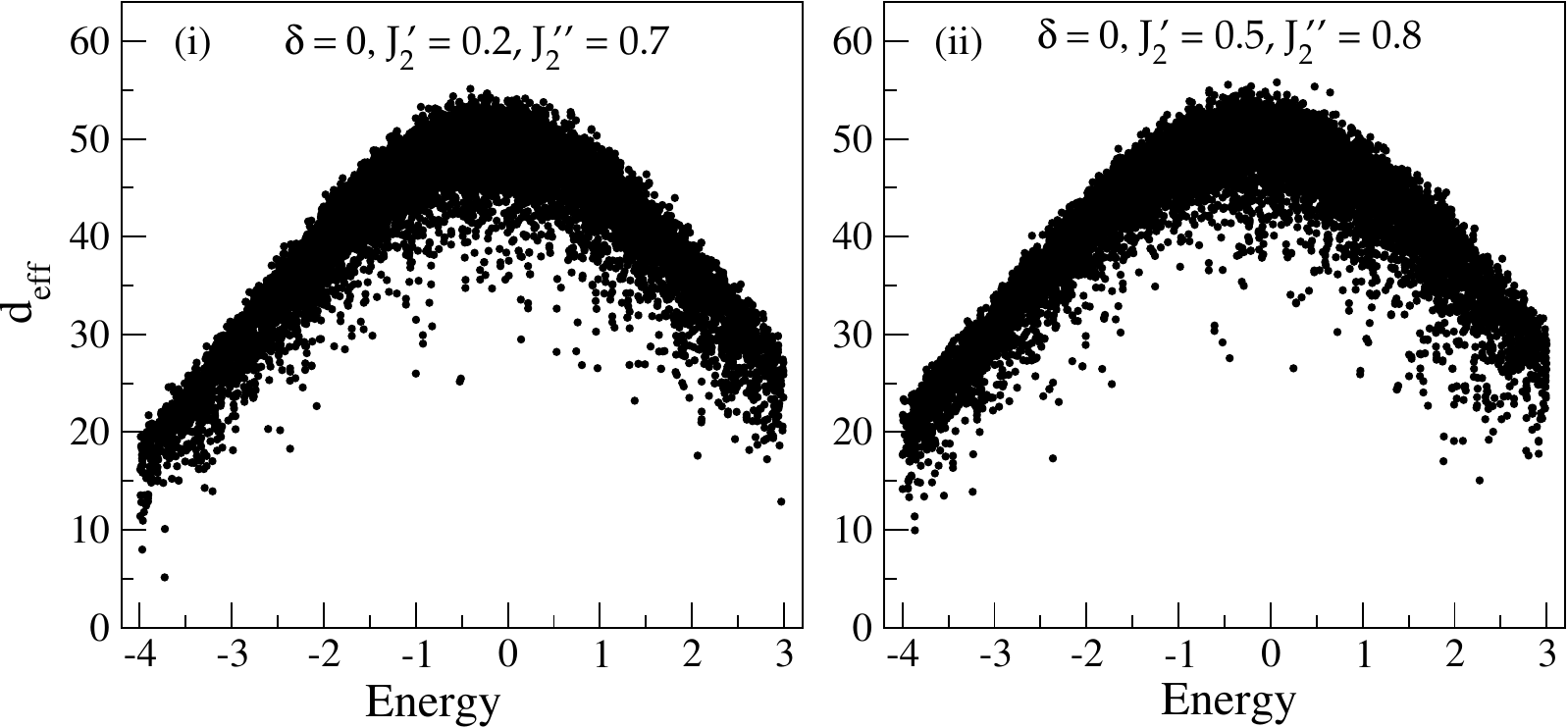}
    \caption{Effective dimension \( d_{\text{eff}} \) profiles with broken SU(2) symmetry ($J_2' \ne J_2''$).}
    \label{deff_nonsym}
\end{figure}

\subsection{Effective dimension of subsystem}
To gain more insight into the thermalization properties of the $J_1-J_2$ model with dimerization, we now study the effective Hilbert space dimension ($d_{\text{eff}}$) of the subsystem for some representative points, ($a$) - ($i$), from Fig. \ref{phase_point}. If $\rho$ represents the reduced density matrix of the subsystem when the full system is in an eigenstate, then the effective Hilbert space dimension of the subsystem is defined as:
\begin{equation}
    d_{\text{eff}} = \frac{1}{\text{Tr}(\rho^2)},
\end{equation}
where Tr represents the trace operation over a basis of the subsystem. We note that this quantity can be considered as an entanglement measure and is related to the (quantum) R\'{e}nyi entropy of order 2: $S_2(\rho)=-\log{\text{Tr}(\rho^2)}=\log({d}_{\text{eff}})$. 
For clarity, we note that when $\rho$ is a pure state, the effective dimension $d_{\text{eff}}=1$, while for a maximally mixed state, $d_{\text{eff}}=D_1$, where $D_1$ denotes the Hilbert space dimension of the subsystem. 

\begin{figure*}
    \centering
    \includegraphics[width=14.6cm, height=10.8cm]{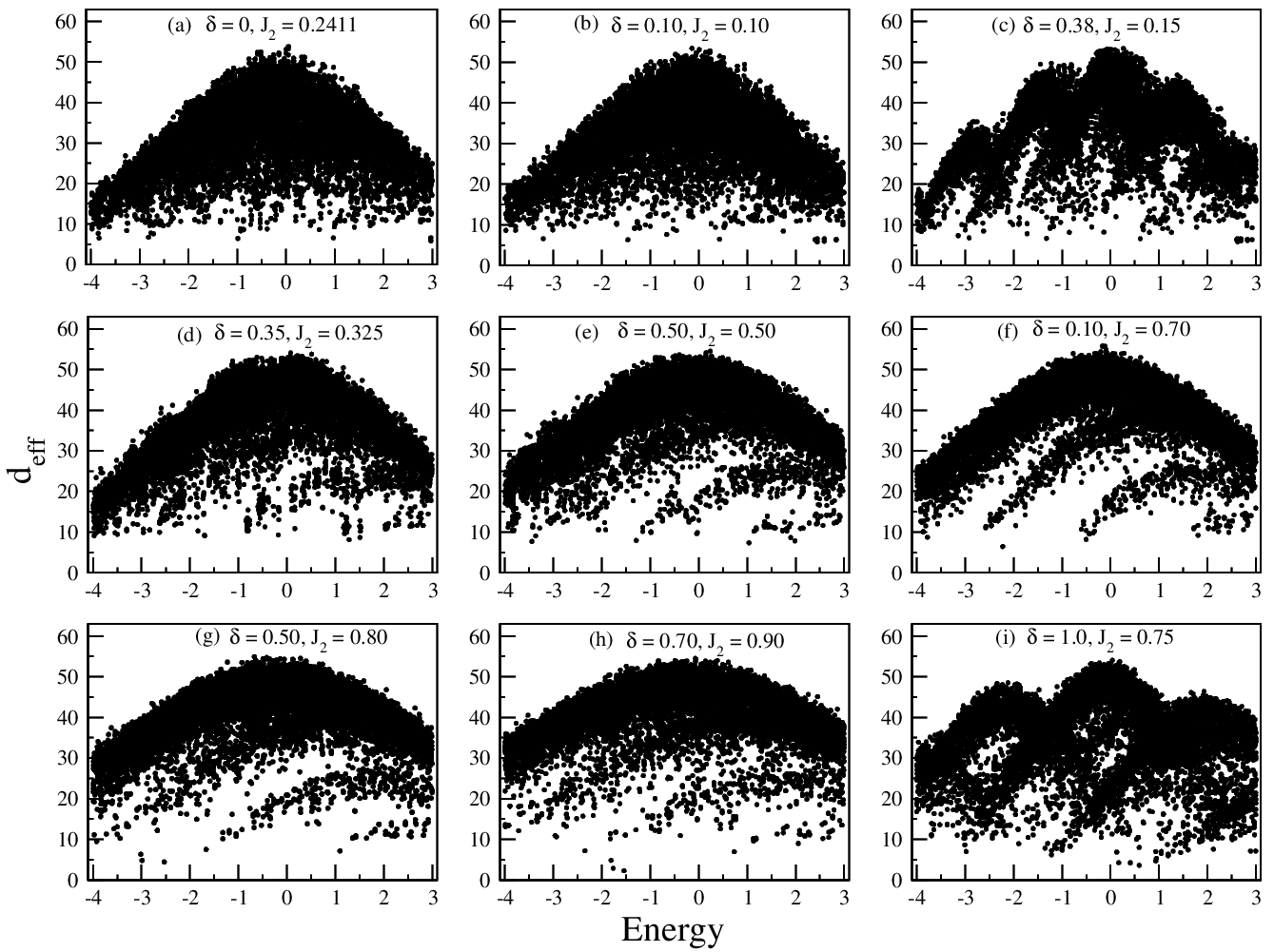}
    \caption{Effective dimension \( d_{\text{eff}} \) of the subsystem  for the individual eigenstates across the energy spectrum is shown for the nine selected points, ($a$) - ($i$), from Fig. \ref{phase_point}.}
    \label{compare_deff}
\end{figure*}

It is known that the average entanglement entropy (or equivalently, $d_{\text{eff}}$) within a narrow energy window reaches its maximum possible value in systems that obey the ETH \cite{Mishra2025}. In such systems, the fluctuations of entropy around the average are also significantly suppressed. These results can be used to understand the thermalization behavior of our spin system. In the following, we first numerically study the individual effects of $\delta$ and $J_2$ on thermalization, subsequently, we investigate the thermalization properties of the spin model at selected points in the parameter space $\delta-J_2$. For the numerical calculations here, we take a system of size $N=16$ with a subsystem of size $N_s=6$.

To understand how dimerization (i.e., nonuniformity in the chain) affects thermalization, we first plot the effective dimension $d_{\text{eff}}$ across the energy spectrum for various eigenstates at selected values of $\delta$, keeping $J_2=0$. The resulting $d_{\text{eff}}$ profiles are shown in Fig. \ref{deff_j2_0}. Likewise, to study the impact of the next-nearest-neighbor interactions, we plot $d_{\text{eff}}$ for different values of $J_2$ with $\delta=0$; these results are presented in Fig.\ref{deff_del_0}. From these figures, we identify three characteristically different types of $d_{\text{eff}}$ profiles:
\begin{enumerate}
    \item A profile with high fluctuations in the values of $d_{\text{eff}}$ without a clear branch structure, as seen for ($\delta=0,J_2=0$) in Fig. \ref{deff_j2_0};
    \item A multi-branched profile where distinct, narrow peaks appear at different energies with comparable prominence, as for ($\delta=0.5,J_2=0$) in the same figure;
    \item A profile with broader, unequal branches, where a weaker structure appears beneath a more dominant one — exemplified by the profile for ($\delta=0,J_2=0.8$) in Fig. \ref{deff_del_0}.
\end{enumerate}

The first type of profile corresponds to a violation of the ETH. The second type of profile results from the grouping of eigenvalues of the system (similar to what we have observed in the DOS profile for the point ($j$) in Fig. \ref{compare_dos}). The third type of profile is understood to be a consequence of the SU(2) symmetry in Hamiltonian of Eq. \ref{eq:hamiltonian}. Each branch in such profile corresponds to different total spin sectors, with the most prominent branch (the upper one) corresponds to the zero total spin. 

To verify that the branches in the third type of $d_{\text{eff}}$ profiles are due to the SU(2) symmetry, we take the same Hamiltonian with a symmetry breaking term:
\begin{align} \label{ham_nonsym}
H &= J_1\sum_{i=1}^{N} \left[ 1 + (-1)^i \delta \right] \mathbf{S}_i \cdot \mathbf{S}_{i+1} \\ \nonumber
&+J_2' \sum_{i=1}^{N} (S_i^x S_{i+2}^x + S_i^y S_{i+2}^y) + J_2'' \sum_{i=1}^{N} S_i^z S_{i+2}^z.
\end{align}
It may be noted that this Hamiltonian is the same as the one appears in Eq. \ref{eq:hamiltonian}, except that now we have different next-nearest-neighbor coupling strengths along the $z$ and $x$ (or $y$) spin components. The $d_{\text{eff}}$ profiles, as appear in Fig. \ref{deff_nonsym}, do not now have branches that appear due to symmetry (as in Fig. \ref{deff_del_0}). Since the original model under study has the SU(2) symmetry, it is expected that the features of the third type of $d_{\text{eff}}$ profile should also be present in the other two types of profiles. But we do not see this clearly since the fluctuations in the first type of profile are very high and the peaks are narrow in the second type of profile.  

Having gained some insight into the effects of having finite $\delta$ and $J_2$ on thermalization, we now study the thermalization behavior of the dimerized $J_1-J_2$ spin chain by analyzing the $d_{\text{eff}}$ profiles at some selected points in the parameter space. The corresponding numerical results are shown in Fig. \ref{compare_deff} (the calculations are done with a system size $N=16$ and subsystem size $N_s=6$). We make the following observations based on the results presented in this figure, and in the figures \ref{deff_j2_0} and \ref{deff_del_0}. When the system lies in the gapless ground-state phase ($0\le J_2 \le J_{2c}, \delta=0$), the ETH is more prone to violation. This result is consistent with the earlier observation that the system thermalizes slowly for small $J_2$ without dimerization \cite{Konstant15}. For larger $J_2$ values and weak dimerization (small $\delta$), the ETH appears to be satisfied within each branch associated with a fixed total spin sector. 

For $J_2=0$, the dimerization ($\delta$) works as integrability breaking parameter. Without frustration, the system is integrable in the two limits: $\delta=0$ and 1. For dimerization in between these two limits, the system is nonintegrable \cite{Konstant16}. It is interesting that for small $\delta$ and $J_2$, although the system is nonintegrable, the thermal phase diagram (Fig. \ref{phase_point}) shows that the fluctuation ($\sigma_{sc}$) in the expectation value of a local operator is reasonably large. Therefore, the system in this regime is expected to show slow thermalization, as observed earlier \cite{Konstant16}.

For intermediate to larger values of $\delta$ (with small $J_2$), the $d_{\text{eff}}$ profile exhibits distinct branches, which arise due to the clustering of eigenvalues or modulations in the density of states (DOS). As mentioned earlier, for large $\delta$ ($\sim 1$) and small $J_2$ ($<0.5$), the system is in a localized phase, hence the study of compliance of ETH is less meaningful here. For an intermediate value of $\delta$ ($\sim 0.5$) and a small value of $J_2$ ($<0.5$), the system is expected to be ETH compliant, although not as strongly as in the regime where $J_2$ is large ($>0.5$).

The results obtained here are overall consistent with what we have seen earlier by studying the expectation value of a local operator. We again see here that the ETH is not obeyed satisfactorily in the limits when both $\delta$ and $J_2$ are small (in the N\'{e}el phase), and the hypothesis holds better in the limits with intermediate $\delta$ values and intermediate to large values of $J_2$ (in the Spiral phase). 

\begin{table}[]
\centering
\setlength{\tabcolsep}{10pt}
\begin{tabular}{ccccc}
\hline
\\[-2.0mm]
Point & ($\delta$, $J_2$)  & $\overline{S}_{vn}$ & $\sigma_{sc}$ \\[1mm]
  \hline
  \hline\\[0.5mm]
        a & (0.00, 0.2411)  & 3.59039 & 0.01698 \\[1mm]
        b & (0.10, 0.10) & 3.61727 & 0.01794 \\[1mm]
        c & (0.38, 0.15) &  3.67362 & 0.01545 \\[1mm]
        d & (0.35, 0.325) & 3.71675 & 0.01266 \\[1mm]
        e & (0.50, 0.50) & 3.73651 & 0.01321 \\[1mm]
        f & (0.10, 0.70) & 3.74837 & 0.01600 \\[1mm]
        g & (0.50, 0.80) & 3.77011 & 0.01143 \\[1mm]
        h & (0.70, 0.90) & 3.76536 & 0.01422 \\[1mm]
        i & (1.0, 0.75) & 3.62846 & 0.02063 \\[1mm]
\hline
\end{tabular}
\caption{The average entropy ($\overline{S}_{vn}$) and the fluctuation in the expectation value of a local operator ($\sigma_{sc}$) are reported here for selected points, ($a$) - ($i$), from Fig. \ref{phase_point}.}
\label{av_vn_entropy}
\end{table}


\subsection{Average VN entropy}
Let $\rho_n$ be the reduced density matrix of the subsystem when the full system is in the eigenket $\ket{n}$. The corresponding von Neumann (VN) entropy of the subsystem is $S_{vn}(\rho_n)=-\text{Tr}(\rho_n\ln{\rho_n})$. We define the average VN entropy over all eigenkets in the following way:
\begin{equation}\label{av_entropy}
    \overline{S}_{vn}=\frac{1}{D}\sum_{n=1}^D S_{vn}(\rho_n),
\end{equation}
where $D$ is the total number of eigenkets (or the dimension of the Hilbert space of the full suystem). It is known that $\overline{S}_{vn}$ follows a volume-law, i.e., this quantity is proportional to the size of the subsystem (in the leading term) \cite{LeBlond2019,Bianchi2022,Vidmar2017}. The volume-law coefficient (or the proportionality constant) takes the maximum possible value for the nonintegrable (chaotic) systems, and is less than the maximum for the integrable systems. In general, we have the following expression for $\overline{S}_{vn}$:
\begin{equation}\label{vol_law}
    \overline{S}_{vn}=\alpha_V N_s + \phi,
\end{equation}
where $\phi$ is a function of both the system size $N$ and the subsystem size $N_s$. It represents the non-universal subleading term, expected to be insignificant compared to the leading term when $1< N_s \ll N$. For a nonintegrable system, in the $S_z=0$ sector, the volume-law coefficient is expected to take a universal value: $\alpha_V = \ln(2) \approx  0.6931 $.    

For selected points in the parameter space $\delta-J_2$, we calculate this average entropy $\overline{S}_{vn}$ to determine to what extent the ETH is satisfied. The numerical results are shown in Table \ref{av_vn_entropy} (the calculations are done with a system size $N = 16$ and subsystem size $N_s = 6$). We also show the corresponding values of $\sigma_{sc}$ to check the consistency of the results. Both results in the table point out the overall same finding about thermalization: ETH holds stronger in the regime where $\delta$ takes intermediate values ($\sim 0.5$) and the values of $J_2$ are between intermediate ($\sim 0.5$) and large ($\sim 1$).  

\begin{figure}[t]
    \centering
    \includegraphics[width=8.2cm, height=6.7cm]{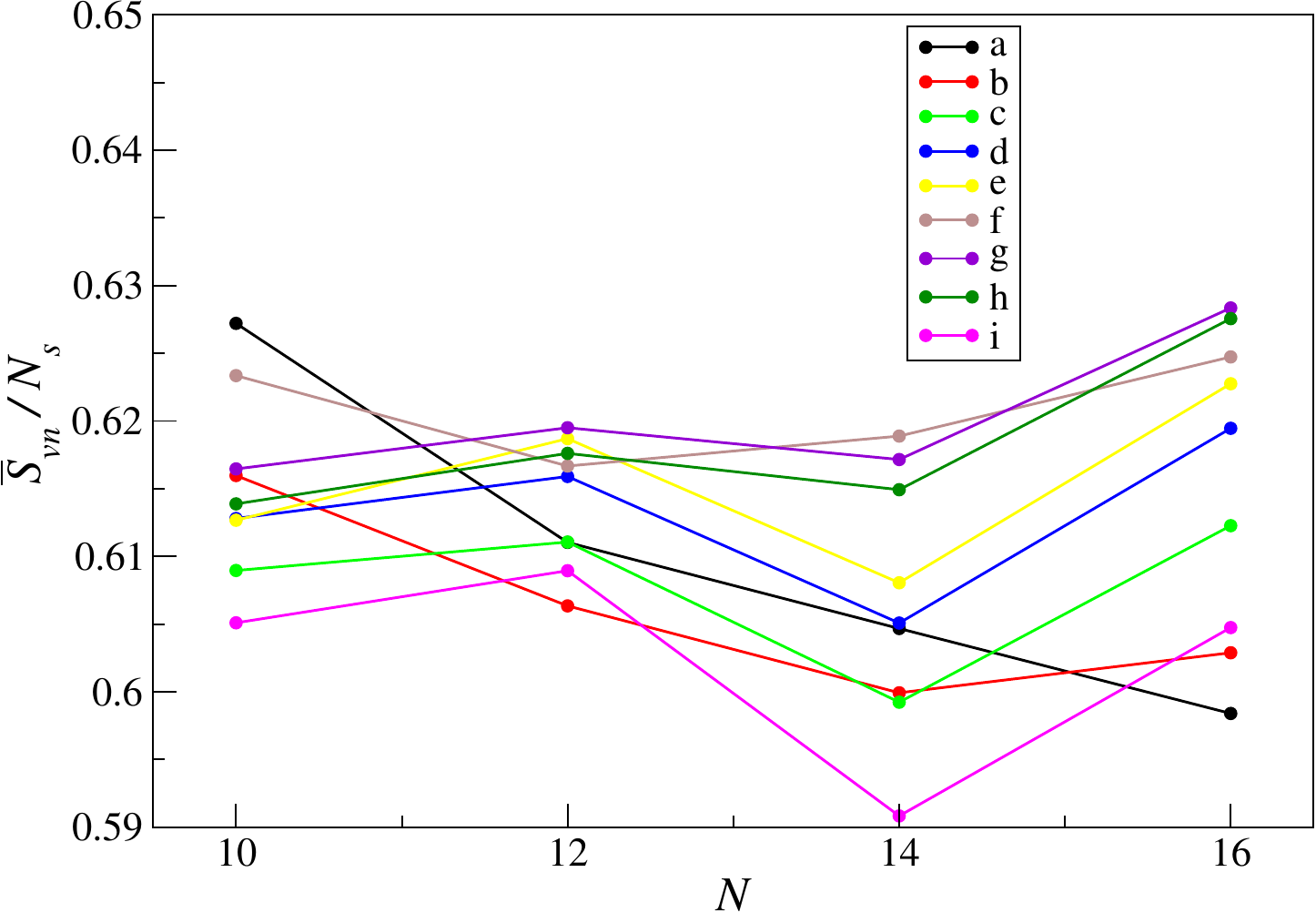}
    \caption{The value of the ratio $\overline{S}_{vn}/N_s$ for different system sizes $N = 8, 10, 12, 14$ and 16 at the selected points $(\delta, J_2)$ from Fig.~\ref{phase_point}.}
    \label{alpha_v}
\end{figure}

\subsection{Finite size scaling}
To better assess whether the system is integrable or nonintegrable at a given point in the $\delta-J_2$ parameter space, one should examine how the relevant physical quantities scale with system size. A good quantity to study here is the ratio $\overline{S}_{vn}/N_s$. In the limit $N_s/N \to 0$, this ratio converges to $\alpha_V$ of Eq. \ref{vol_law}. With increasing system size $N$ (keeping $N_s/N$ fixed at a small value), $\overline{S}_{vn}/N_s$ is expected to increase and converge to $\alpha_V=\ln(2)$ if the system is nonintegrable (chaotic). On the other hand, the ratio should take a lower than the maximum value ($\alpha_V<\ln(2)$) for a large system size if the system is integrale. 

In Fig. \ref{alpha_v}, we plot the ratio $\overline{S}_{vn}/N_s$ as a function of the system size $N$ for selected points in the parameter space.  The calculations are done for the system sizes $N=8, 10, 12, 14$ and $16$ (the largest system size we could access) with corresponding subsystem sizes $N_s=2,3,4,5$ and $6$. We choose the particular values of $N_s$ so that the ratio $N_s/N$ does not change much as we increase the system size. We see from the figure that, after some smaller values of $N$, the ratio $\overline{S}_{vn}/N_s$ increases (especially from $N=14$ to 16) for all points except for the point associated with the gapless ground-state phase. 

This finite-size scaling behavior reinforces our earlier observation: ETH is more likely to be violated by the system having a gapless ground-state phase, while it tends to hold for the system with the gapped ground-state phases. Moreover for finite but large system, ETH appears to be obeyed more robustly in the region where $\delta$ is of intermediate strength and $J_2$ takes moderate to large values.

\section{Conclusion} \label{sec5}
The thermalization property of the dimerized $J_1-J_2$ spin-1/2 model is investigated here. This model offers a minimal yet nontrivial setting to study thermalization under both inhomogeneity (via dimerization) and extended-range interactions (via next nearest-neighbor couplings). Our results suggest that the system is more likely to violate ETH when its ground-state phase is gapless ($0\le J_2\le J_{2c}$ with $\delta =0$). Although the system is seen to obey ETH when its ground-state phases are gapped, a stronger compliance is expected in the region where $\delta$ is of intermediate strength and $J_2$ takes moderate to large values (the region is associated with the spiral ground-state phase of the model).

These findings demonstrate that controlled bond alternation provides a powerful tool for engineering thermalization in quantum simulators, with potential applications in quantum memory design. 
The results highlight how subtle modifications to interaction patterns can dramatically alter many-body dynamics, suggesting new pathways for controlling quantum thermalization in programmable quantum devices. 

\begin{acknowledgments}
SS thanks Ranjan Modak for useful discussions. 
\end{acknowledgments}

\bibliography{apssamp}

\end{document}